\documentclass[12pt]{iopart}

\usepackage{graphicx}

\usepackage{cite}

\usepackage{dcolumn}

\begin{document}

\title[Asymptotic Iteration Method]
{Application of the Asymptotic Iteration Method to a Perturbed Coulomb
Model}

\author{Paolo Amore\dag \ and Francisco M Fern\'{a}ndez\ddag}

\address{\dag\ Facultad de Ciencias, Universidad de Colima,
Bernal D\'{i}az del Castillo 340, Colima, Colima, Mexico.}

\ead{paolo@ucol.mx}

\address{\ddag\ INIFTA (Conicet, UNLP), Blvd. 113 y 64 S/N, Sucursal 4,
Casilla de Correo 16, 1900 La Plata, Argentina}

\ead{fernande@quimica.unlp.edu.ar}

\date{\today}

\begin{abstract}
We show that the asymptotic iteration method converges and yields accurate
energies for a perturbed Coulomb model. We also discuss alternative
perturbation approaches to that model.
\end{abstract}

\maketitle

\section{Introduction \label{sec:intro}}

The asymptotic iteration method (AIM) is an iterative algorithm for the
solution of Sturm--Liouville equations~\cite{CHS03,F04}. Although this
method does not seem to be better than other existing approaches, it has
been applied to quantum--mechanical~\cite{B05,B06,CHS05} as well as
mathematical problems~\cite{B05b}. For example, the AIM has proved suitable
for obtaining both accurate approximate and exact eigenvalues~\cite
{CHS03,F04,B05,B05b,B06,CHS05} and it has also been applied to the
calculation of Rayleigh--Schr\"{o}dinger perturbation coefficients~\cite
{B06,CHS05b}.

Recently, Barakat applied the AIM to a Coulomb potential with a radial
polynomial perturbation~\cite{B06}. By means of a well--known transformation
he converted the perturbed Coulomb problem into an anharmonic oscillator.
Since straightforward application of the AIM exhibited considerable
oscillations and did not appear to converge Barakat resorted to perturbation
theory in order to obtain acceptable results \cite{B06}.

It is most surprising that the straightforward application of the AIM failed
for the anharmonic oscillator studied by Barakat~\cite{B06} since it had
been found earlier that the approach should be accurate in such cases~\cite
{F04}.

The main purpose of this paper is to verify whether the AIM gives accurate
eigenvalues of the perturbed Coulomb model or if its sequences are
oscillatory divergent as mentioned above. We also discuss the application of
perturbation theory to that model.

In Sec.~\ref{sec:model} we present the model and discuss useful scaling
relations for the potential parameters. In Sec.~\ref{sec:AIM} we apply the
AIM to the perturbed Coulomb model directly; that is to say we do not
convert it into an anharmonic oscillator. In Sec.~\ref{sec:PT} we outline
alternative perturbation approaches, and in Sec.~\ref{sec:concl} we
interpret our results and draw conclusions.

\section{The model \label{sec:model}}

The problem studied by Barakat~\cite{B06} is given by the following radial
Schr\"{o}dinger equation
\begin{eqnarray}
\hat{H}\Psi &=&E\Psi ,  \nonumber \\
\hat{H} &=&-\frac{1}{2}\frac{d^{2}}{dr^{2}}+\frac{l(l+1)}{2r^{2}}+V(r)
\nonumber \\
V(r) &=&-\frac{Z}{r}+gr+\lambda r^{2},  \label{eq:Schro}
\end{eqnarray}
where $l=0,1,2,\ldots $ is the angular--momentum quantum number, and the
boundary conditions are $\Psi (0)=\Psi (\infty )=0$. We restrict to the case
$\lambda >0$ in order to have only bound states; on the other hand, $Z$ and $%
g$ can take any finite real value.

It is most useful to take into account the scaling relations
\begin{eqnarray}
E(Z,g,\lambda ) &=&Z^{2}E(1,gZ^{-3},\lambda
Z^{-4})=|g|^{2/3}E(Z|g|^{-1/3},g|g|^{-1},\lambda |g|^{-4/3})  \nonumber \\
&=&\lambda ^{1/2}E(Z\lambda ^{-1/4},g\lambda ^{-3/4},1).  \label{eq:scaling}
\end{eqnarray}
Notice that we can set either $Z$ or $\lambda $ equal to unity without loss
of generality, and that, for example, $E(1,-g,\lambda )=E(-1,g,\lambda )$.
Following Barakat~\cite{B06} we choose $n=0,1,\ldots $ to be the radial
quantum number, and we may define a ``principal '' quantum number $\nu
=n+l+1=1,2,\ldots $.

\section{Direct application of the AIM\label{sec:AIM}}

Barakat mentions that straightforward application of the AIM does not give
reasonable results because the sequences oscillate when the number of
iteration is greater than $30$ approximately~\cite{B06}. This conclusion is
surprising because it has been shown that the AIM yields accurate results
for anharmonic oscillators~\cite{F04}, and Barakat converted the perturbed
Coulomb model into one of them\cite{B06}. In this section we apply the AIM
directly to the original radial Schr\"{o}dinger equation (\ref{eq:Schro}).

By means of the transformation $\psi (r)=\phi (r)y(r)$ we convert the
perturbed Coulomb model (\ref{eq:Schro}) into a Sturm--Liouville equation
for $y(r)$:
\begin{eqnarray}
y^{\prime \prime }(r) &=&Q(r)y^{\prime }(r)+R(r)y(r)  \nonumber \\
Q(r) &=&-\frac{2\phi ^{\prime }(r)}{\phi (r)}  \nonumber \\
R(r) &=&\left\{ 2[V(r)-E]-\frac{\phi ^{\prime \prime }(r)}{\phi (r)}\right\}
,  \label{eq:St_Li_gen}
\end{eqnarray}
where $\phi (r)$ is arbitrary. It seems reasonable to choose
\begin{equation}
\phi (r)=r^{l+1}e^{-\beta r-\alpha r^{2}}  \label{eq:phi(r)}
\end{equation}
that resembles the asymptotic behaviour of the eigenfunction for a harmonic
oscillator when $\beta =0$ or for a Coulomb interaction when $\alpha =0$. It
leads to
\begin{eqnarray}
Q(r) &=&4\alpha r-\frac{2(l+1)}{r}+2\beta  \nonumber \\
R(r) &=&\left( 2\lambda -4\alpha ^{2}\right) r^{2}+(2g-4\alpha \beta )r+%
\frac{2\beta (l+1)-2Z}{r}  \nonumber \\
&&+2\alpha (2l+3)-2E-\beta ^{2}.  \label{eq:P(r)_Q(r)}
\end{eqnarray}
We can set the values of the two free parameters $\alpha $ and $\beta $ to
obtain the greatest rate of convergence of the AIM sequences. From now on we
call asymptotic values of $\alpha $ and $\beta $ to such values of those
parameters that remove the terms of $R(r)$ that dominate at large $r$; that
is to say: $\beta =g/(2\alpha )$ and $\alpha =\sqrt{\lambda /2}$. Since the
asymptotic values of the free parameters do not necessarily lead to the
greatest convergence rate\cite{F04}, in what follows we will also look for
optimal values of $\alpha $.

The Sturm--Liouville equation (\ref{eq:St_Li_gen}) with the functions $Q(r)$
and $R(r)$ (\ref{eq:P(r)_Q(r)}) is suitable for the application of the AIM.
We do not show the AIM equations here because they have been developed and
discussed elsewhere\cite{CHS03,F04}. Since the AIM quantization condition
depends not only on the energy but also on the variable $r$ for non--exactly
solvable problems, we have to choose a convenient value for the latter\cite
{CHS03,F04}. Later on we will discuss the effect of the value of $r$ on the
convergence of the method; for the time being we follow Barakat~\cite{B06}
and select the positive root of $\phi ^{\prime }(r)=0$:
\begin{equation}
r_{0}=\frac{\sqrt{8\alpha (l+1)+\beta ^{2}}-\beta }{4\alpha }  \label{eq:r_0}
\end{equation}

For concreteness we restrict to $Z=\lambda =1$ and $n=l=0$, and select $%
g=-2,-1,1,2$ from Barakat's paper\cite{B06}. As expected from earlier
calculations on anharmonic oscillators~\cite{F04}, the rate of convergence
of the AIM depends on the value of $\alpha $. In order to investigate this
point we choose $g=-2$ because it is the most difficult of all the cases
considered here. More precisely, we focus on the behaviour of the
logarithmic error $L_{N}=\log |E^{(N)}-E^{exact}|$, where $E^{(N)}$ is the
AIM energy at iteration $N$ and $E^{exact}=-1.1716735847196510437987056$ was
obtained by means of the rapidly converging Riccati--Pad\'{e} Method (RPM)~%
\cite{F95,F96} from sequences of determinants of dimension $D=2$ through $%
D=22$.

We first consider the asymptotic value $\alpha =1/\sqrt{2}$.
Fig.~\ref
{fig:asympt} shows that $L_{N}$ decreases rapidly with $N$ when $%
N<\approx 20$ and then more slowly but more smoothly for $N>20$.
In the transition region about $N\approx 20$ we observe
oscillations that can mislead one into believing that the AIM
starts to diverge.

\begin{figure}[tbp]
\begin{center}
\includegraphics[width=9cm]{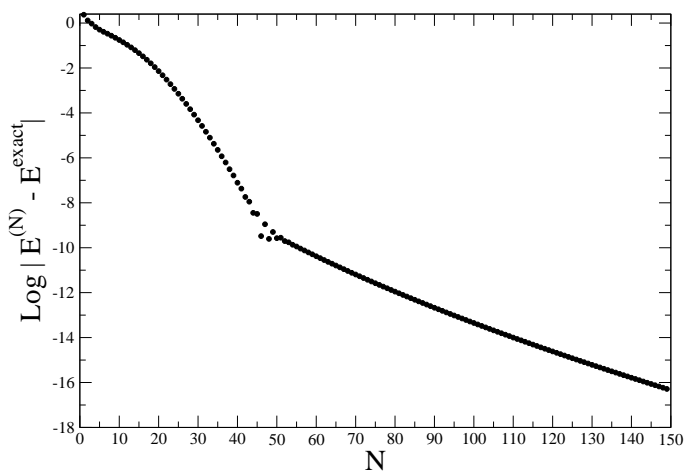}
\end{center}
\caption{Logarithmic error for the energy for $g=-2$ as a function of $N$ for an almost optimal value $%
\alpha=1/2$}
\label{fig:optimal}
\end{figure}

Fig.~\ref{fig:optimal} shows that the behaviour of $L_{N}$ for a
nearly optimal value $\alpha =1/2$ is similar to the previous
case, except that the transition takes place at a larger value of
$N$ and the convergence rate is
greater. More precisely, $L_{N}$ decreases rapidly with $N$ when $%
N<\approx 50$ approximately as $L_{N}\approx
0.22-0.064N-0.0029N^{2}$ and more slowly and smoothly for $N>50$
as $L_{N}\approx -6.5-0.068N$. Again, the transition region
exhibits oscillations.

Table~\ref{tab:Table2} shows the ground--state energies for
$g=-2,-1,1,2$ and the corresponding nearly optimal values of
$\alpha $. We estimated those eigenvalues from the sequences of
AIM roots for $N=10$ through $N=80$. Notice that the optimal
values of $\alpha $ in Table~\ref{tab:Table2} depend on $g$ and do
not agree with the asymptotic value $\alpha =\sqrt{1/2}$.
Table~\ref{tab:Table2} also shows that the AIM eigenvalues agree
with those calculated by means of the RPM~\cite{F95,F96} from
sequences of determinants of dimension $D=2$ through $D=15$.

The rate of convergence also depends on the chosen value of $r$. The
calculation of $L_{N}$ as a function of $\xi =r/r_{0}$ shows that $L_{N}(\xi
)$ exhibits a minimum at $\xi _{N}$ and that $\xi _{N}$ increases with $N$
approximately as $\xi _{N}=0.435+0.005N$ (for $g=-2$). However, in order to
keep the application of the AIM as simple as possible we just choose $%
r=r_{0} $ for all the calculations.

\section{Alternative perturbation approaches\label{sec:PT}}

Barakat~\cite{B06} first converted the radial Schr\"{o}dinger equation (\ref
{eq:Schro}) into another one for an anharmonic oscillator by means of the
standard transformations $r=u^{2}$ and $\Phi (u)=u^{-1/2}\Psi (u^{2})$.
Finally, he derived the Sturm--Liouville problem
\begin{equation}
f^{^{\prime \prime }}(u)+2\left( \frac{L+1}{u}-\alpha u^{3}\right) f^{\prime
}(u)+\left( \epsilon u^{2}-8gu^{4}+8Z\right) f(u)=0,  \label{eq:SL_u}
\end{equation}
where $L=2l+1/2$ and $\epsilon =8E-(2L+5)\alpha $, through factorization of
the asymptotic behaviour of the solution:
\begin{equation}
\Phi (u)=u^{L+1}e^{-\alpha u^{2}/4}f(u),\;\alpha =\sqrt{8\lambda }.
\label{eq:factor1}
\end{equation}
Notice that present $\alpha $ and Barakat's $\alpha $ are not exactly the
same but they have a close meaning and are related by $\alpha
_{asymptotic}^{present}=\alpha _{Barakat}/4$. Since Barakat's application of
the AIM to Eq. (\ref{eq:SL_u}) did not appear to converge~\cite{B06} he
opted for a perturbation approach that consists of rewriting Eq. (\ref
{eq:SL_u}) as
\begin{equation}
f^{^{\prime \prime }}(u)+2\left( \frac{L+1}{u}-\alpha u^{3}\right) f^{\prime
}(u)+\left[ \epsilon u^{2}+\gamma \left( -8gu^{4}+8Z\right) \right] f(u)=0
\label{eq:SL_u_PT}
\end{equation}
and expanding the solutions in powers of $\gamma $:
\begin{equation}
f(u)=\sum_{j=0}^{\infty }f^{(j)}(u)\gamma ^{j},\;\epsilon
=\sum_{j=0}^{\infty }\epsilon ^{(j)}\gamma ^{j}  \label{eq:AIM_PT_series_u}
\end{equation}
The perturbation parameter $\gamma $ is set equal to unity at the end of the
calculation. The series for the energy exhibits considerable convergence
rate and consequently Barakat obtained quite accurate results with just two
to five perturbation corrections~\cite{B06}. Barakat calculated the
coefficient $\epsilon ^{(0)}$ exactly and all the others approximately~\cite
{B06}.

The model (\ref{eq:Schro}) is suitable for several alternative
implementations of perturbation theory in which we simply write $%
V(r)=V_{0}(r)+\gamma V_{1}(r)$ and expand the solutions in powers of $\gamma
$.

If we choose $V_{0}(r)=-Z/r$ (when $Z>0$) and $V_{1}(r)=gr+\lambda r^{2}$
then we can calculate all the perturbation coefficients exactly by means of
well known algorithms~\cite{F00}. One easily realizes that the perturbation
series can be rearranged as
\begin{equation}
E=Z^{2}\sum_{i=0}^{\infty }\sum_{j=0}^{\infty }d_{ij}g^{i}\lambda
^{j}Z^{-(3i+4j)}.  \label{eq:E_series_Z}
\end{equation}
It is well known that this series is asymptotic divergent for all values of
the potential parameters.

The other reasonable perturbation split of the potential energy is $%
V_{0}(r)=\lambda r^{2}$, $V_{1}(r)=-Z/r+gr$. In this case we can rearrange
the series as
\begin{equation}
E=\lambda ^{1/2}\sum_{i=0}^{\infty }\sum_{j=0}^{\infty
}c_{ij}Z^{i}g^{j}\lambda ^{-(i+3j)/4}.  \label{eq:E_series_lambda}
\end{equation}
One expects that this series has a finite radius of convergence. This is
exactly the series obtained by Barakat~\cite{B06} by means of the AIM and,
consequently, it is not surprising that he derived accurate results from it.
In this case one can obtain exact perturbation corrections at least for the
first two energy coefficients.

For simplicity we concentrate on the states with $n=0$. The eigenfunctions
and eigenvalues of order zero are
\begin{eqnarray}
\Psi _{0l}^{(0)}(r) &=&\frac{\sqrt{2}(2\lambda )^{(2l+3)/8}}{\Gamma (l+3/2)}%
r^{l+1}e^{-\sqrt{2\lambda }r^{2}/2}  \nonumber \\
E_{0l}^{(0)} &=&\frac{\sqrt{2\lambda }}{2}(2l+3)  \label{eq:E^(0)}
\end{eqnarray}
respectively. With the unperturbed eigenfunctions one easily obtains the
perturbation correction of first order to the energy
\begin{equation}
E_{0l}^{(1)}=\frac{(l+1)!g}{(2\lambda )^{1/4}\Gamma (l+3/2)}-\frac{%
l!Z(2\lambda )^{1/4}}{\Gamma (l+3/2)}  \label{eq:E^(1)}
\end{equation}
that is the term of the series (\ref{eq:E_series_lambda}) with $i+j=1$. One
can easily carry out the same calculation for the states with $n>0$ using
the appropriate eigenfunctions of the harmonic oscillator.

Equation (\ref{eq:E^(1)}) yields all the numerical results for $\epsilon
_{0l}^{(1)}$ in Tables 1-3 of Barakat's paper~\cite{B06}. In particular, $%
E_{0l}^{(1)}=0$ when $g=\sqrt{2\lambda }Z/(l+1)$ as in Table 1 of Barakat's
paper~\cite{B06}. This particular relationship between the potential
parameters also leads to exact solutions of the eigenvalue equation (\ref
{eq:Schro}). Some of them are given by
\begin{eqnarray}
\Psi _{0l}^{exact}(r) &=&N_{l}r^{l+1}e^{-\alpha r^{2}-\beta r},\;\alpha =%
\sqrt{\frac{\lambda }{2}},\;\beta =\frac{Z}{l+1},  \nonumber \\
E_{0l}^{exact} &=&\alpha (2l+3)-\frac{Z^{2}}{2(l+1)^{2}},\;g=\frac{\sqrt{%
2\lambda }Z}{l+1},  \label{eq:QES}
\end{eqnarray}
where $N_{l}$ is a normalization constant.

\section{Conclusions \label{sec:concl}}

We have shown that the AIM converges for the perturbed Coulomb
model if the values of the free parameters in the factor function
that converts the Schr\"{o}dinger equation into a Sturm--Liouville
one are not too far from optimal. It is clear that it is not
necessary to transform the perturbed Coulomb model into an
anharmonic oscillator for a successful application of the AIM. Our
results do not exhibit the oscillatory divergence reported by
Barakat\cite{B06} even when choosing the asymptotic value of
$\alpha$.

The perturbation approach proposed by Barakat~\cite{B06} is equivalent to
choosing the harmonic oscillator as unperturbed or reference Hamiltonian,
and if we apply perturbation theory to the original radial Schr\"{o}dinger
equation we easily obtain two energy coefficients exactly instead of just
only one. It is worth mentioning that the coefficients calculated by Barakat~%
\cite{B06} are quite accurate and, consequently, the resulting series
provide a suitable approach for the eigenvalues of the perturbed Coulomb
potential. This application of the AIM to perturbation theory is certainly
much more practical than the calculation of exact perturbation corrections
proposed earlier~\cite{CHS05b} that can certainly be carried out more
efficiently by other approaches~\cite{F00}.

\noindent \textbf{Acknowledgements}

\noindent P.A. acknowledges support from Conacyt grant
C01-40633/A-1

\begin{table}[tbp]
\caption{Energies for some $g$ values calculated by the AIM with
$N=80$ and RPM~\protect\cite{F95,F96} with $D=15$}
\label{tab:Table2}
\begin{center}
\par
\begin{tabular}{rrD{.}{.}{18}D{.}{.}{17}}

$g$ &  \multicolumn{1}{c}{$\alpha$} & \multicolumn{1}{c}{AIM} &
\multicolumn{1}{c}{RPM} \\ \hline

-2 & 0.5& -1.17167358472& -1.1716735847197\\
-1 & 0.3& -0.226186875190871929& -0.2261868751908719\\
1 & 0.3& 1.33284549226484083&1.3328454922648408349 \\
 2&0.5 &2.014906226463 & 2.0149062264617370560\\

\end{tabular}
\end{center}
\end{table}

\begin{figure}[tbp]
\begin{center}
\includegraphics[width=9cm]{Fig_error_barakat.eps}
\end{center}
\caption{Logarithmic error for the energy for $g=-2$ as a function of $N$ for $\alpha=1/\protect\sqrt{2%
}$}
\label{fig:asympt}
\end{figure}

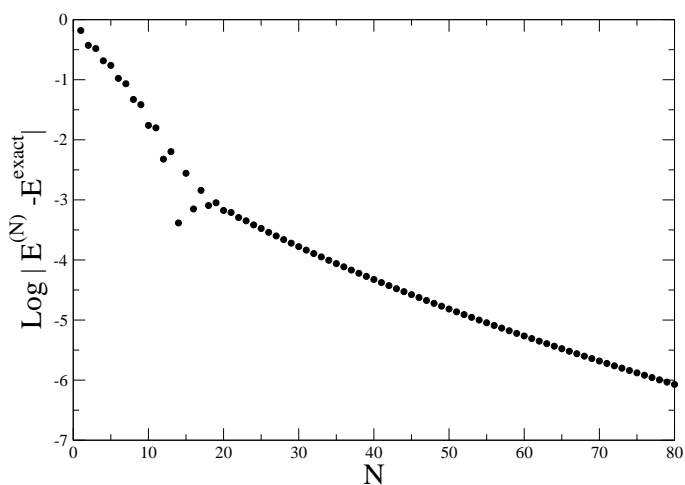
\end{document}